\documentclass[]{spie}  

\usepackage{amsmath,amsfonts,amssymb}
\usepackage{graphicx}
\usepackage[colorlinks=true, allcolors=blue]{hyperref}
\usepackage{xspace}

\title{Reducing the Athena WFI charged particle background: Results from Geant4 simulations}

\author[a]{Catherine E. Grant}
\author[a]{Eric D. Miller}
\author[a]{Marshall W. Bautz}
\author[b]{Tanja Eraerds}
\author[c]{Silvano Molendi}
\author[d]{Jonathan Keelan}
\author[d]{David Hall}
\author[d]{Andrew D. Holland}
\author[e]{Ralph P. Kraft}
\author[e]{Esra Bulbul}
\author[e]{Paul Nulsen}
\author[f]{Steven Allen}
\affil[a]{MIT Kavli Institute for Astrophysics \& Space Research, Cambridge, Massachusetts, USA}
\affil[b]{Max Planck Institute for Extraterrestrial Physics, Garching, Germany}
\affil[c]{INAF/IASF-Milano, Milano, Italy}
\affil[d]{Centre for Electronic Imaging, The Open University, Milton Keynes, UK}
\affil[e]{Center for Astrophysics $|$ Harvard \& Smithsonian, Cambridge, Massachusetts, USA}
\affil[f]{Department of Physics, Stanford University, Stanford, California, USA}

\authorinfo{Further author information: (Send correspondence to C.E.G.)\\C.E.G.: E-mail: cgrant@mit.edu}

\def\chandra  {\textit{Chandra\/}\xspace}
\def\xmm      {\textit{XMM-Newton\/}\xspace}
\def\swift      {\textit{Swift\/}\xspace}
\def\pcor    {$P_{cor}$}

\begin{document} 
\maketitle

\begin{abstract}
One of the science goals of the Wide Field Imager (WFI) on ESA’s Athena X-ray observatory is to map hot gas structures in the universe, such as clusters and groups of galaxies and the intergalactic medium. These deep observations of faint diffuse sources require low background and the best possible knowledge of that background. The WFI Background Working Group is approaching this problem from a variety of directions. Here we present analysis of Geant4 simulations of cosmic ray particles interacting with the structures aboard Athena, producing signal in the WFI. We search for phenomenological correlations between these particle tracks and detected events that would otherwise be categorized as X-rays, and explore ways to exploit these correlations to flag or reject such events in ground processing.  In addition to reducing the Athena WFI instrumental background, these results are applicable to understanding the particle component in any silicon-based X-ray detector in space.
\end{abstract}


\section{INTRODUCTION}
\label{sec:intro}
Athena (Advanced Telescope for High ENergy Astrophysics), ESA's next large X-ray observatory, is scheduled to launch in the early 2030s to Earth-Sun L1 or L2\cite{athena}.  Athena will explore the hot and energetic universe using two selectable focal plane instruments.  The X-ray Integral Field Unit (X-IFU) will provide spatially resolved high resolution spectroscopy\cite{xifu} and the Wide Field Imager (WFI) will provide moderate spectroscopy over a large 40~arcminute field of view\cite{wfi}.  The WFI makes use of DEPFET (depleted p-channel field-effect transistor) active pixel sensor arrays with a pixel size of 130~$\mu$m $\times$ 130~$\mu$m that are fully depleted to 450~$\mu$m.  The energy range is 0.2--15~keV with a full frame readout time of 5~msec.

As some of the Athena science goals require study of faint diffuse emission, such as that in clusters of galaxies, there has been a substantial effort to better understand, model, and reduce the instrumental background produced by particles and high energy photons, both for the observatory as a whole and for the WFI in particular.  The Athena science requirement for the unfocused non-X-ray background is $< 5.5 \times 10^{-3}$~counts/s/cm$^2$/keV in the 2--7~keV band and knowledge of the background to within a few percent.  The background reduction studies include shielding and coating choices, as well as software mitigation, both on-board and on the ground.  A summary of these ongoing efforts by the WFI team and the WFI Background Working Group (BWG) is given in Ref.~\citenum{wfibkg}.  

 A large effort has been underway for several years to predict and model the expected WFI particle background using Geant4\cite{Geant4} simulations, and to use these simulations to inform the design of both the camera shielding and event filtering on-board and in ground processing.\cite{wfibkg}  In this work, we use a set of these Geant4 simulations of cosmic rays interacting with the WFI camera body to model the expected unrejected particle background and explore techniques to separate this signal from the desired X-ray signal produced by celestial objects under observation. In particular, we study correlations between those unrejected events and cosmic ray tracks produced by the same primary particle interaction. In this paper we present highlights of our analysis of Geant4 simulations of cosmic ray particles interacting with the structures on board Athena, which produce signal in the WFI, including efforts to effectively filter background induced events and to improve knowledge of the particle background. These results will be applicable to understanding the particle component in any silicon-based X-ray detector in space. 
 
 This proceedings paper is excerpted from a much more comprehensive work that is in preparation.\cite{Miller20}. In Section~\ref{sect:geant4}, we describe the Geant4 simulation output and in Section~\ref{sect:events}, how this output was converted into simulated WFI frames and event lists. Section~\ref{sect:analysis} describes some characteristics of the simulated background signal and Section~\ref{sect:sum} provides a summary of these results.

\section{GEANT4 SIMULATIONS}
\label{sect:geant4}

The Geant4 simulations were performed at Open University and consisted of 133 runs of 1,000,000 Galactic cosmic ray (GCR) proton primaries per run, drawn from the CREME 96 standard spectral model for solar minimum \cite{Tylka1997} and generated on a 70-cm radius sphere surrounding the WFI instrument. These simulations used a simplified WFI mass model designated E0015261, which includes the camera, proton shield, filter wheel, and baffle, but excludes a graded-Z shield under later study by the WFI BWG to reduce the impact of secondary electrons produced by GCR interactions in the proton shield. This is the same mass model used to obtain results previously presented,\cite{wfibkg} and we refer the reader there for more detailed information about the Geant4 simulation setup and operation. For each GCR primary that generated signal charge in the WFI detector, the data include the deposited energy in keV in each pixel and information about the particle (primary or secondary) responsible for the deposition.  The vast majority of simulated primaries do not interact with the WFI detector; indeed, only 936,934 of 133,000,000 (0.7\%) produce signal in any pixels. 

The Geant4 output was structured into two different formats for further analysis.  The first dataset was structured on a primary-by-primary basis, hereafter referred to as ``single-primary'' frames, and this was used to explore fundamental properties of the signal produced by individual cosmic rays and search for useful correlations between particle tracks and events that look like X-rays that could be exploited to flag the latter. The second type of dataset has primary GCRs randomly sorted into frames of a given frame time, 5~ms or 2~ms, to simulate the real-world WFI background. Two frame times were studied to explore the effects of faster readout on background reduction. Analyses of these finite-frame datasets were compared to the Geant4 analysis by other team members \cite{wfibkg} and to test possible background-reduction algorithms on realistic WFI observations.

The proton flux from these simulations produces an average 2--7 keV unrejected count rate consistent with that derived previously by the WFI BWG for protons only, $5\times10^{-3}$ cm$^{-2}$\,s$^{-1}$\,keV$^{-1}$ \cite{wfibkg}. However, since the real particle background environment includes other species such as GCR alpha particles, electrons, and gamma rays, we increased the proton flux by 40\% to account for these primaries missing from the Geant4 simulations. This produced a total average 2--7 keV unrejected count rate consistent with that found by previous Geant4 analysis amongst the BWG\cite{wfibkg}, $\sim7\times10^{-3}$ cm$^{-2}$\,s$^{-1}$\,keV$^{-1}$. We note that the details of the secondary interactions are likely different between protons and these other species, but to first order this is a reasonable approximation. 

The scaled GCR primary flux yields a total effective exposure time of 1505~sec for the 133 million primaries, a rate of $8.84\times10^{4}$~s$^{-1}$, or 441.9 per 5-ms frame (176.8 per 2-ms frame). Using this as the mean rate, each of the 133 million primaries was assigned a random arrival time drawn from an exponential distribution, appropriate for modeling arrival intervals of this Poisson process. Primaries were then assigned into each frame according to the order in which they were simulated.  We determine a mean rate of 3.11 interacting primaries per frame in the 300,967 5-ms frames that were simulated. Of these frames, 95.5\% have signal in them, consistent with the expectation from the assumed Poisson distribution.

For each case (single-primary, 5-ms, and 2-ms frames), each frame with signal was turned into an image of pixel values using the pixel X, Y, and deposited energy information provided by Geant4. Further details of the simulations can be found in Ref.~\citenum{Miller20}.

\section{Identifying events and particle tracks}
\label{sect:events}

\begin{figure}[t]
\begin{center}
\includegraphics[width=3.0in]{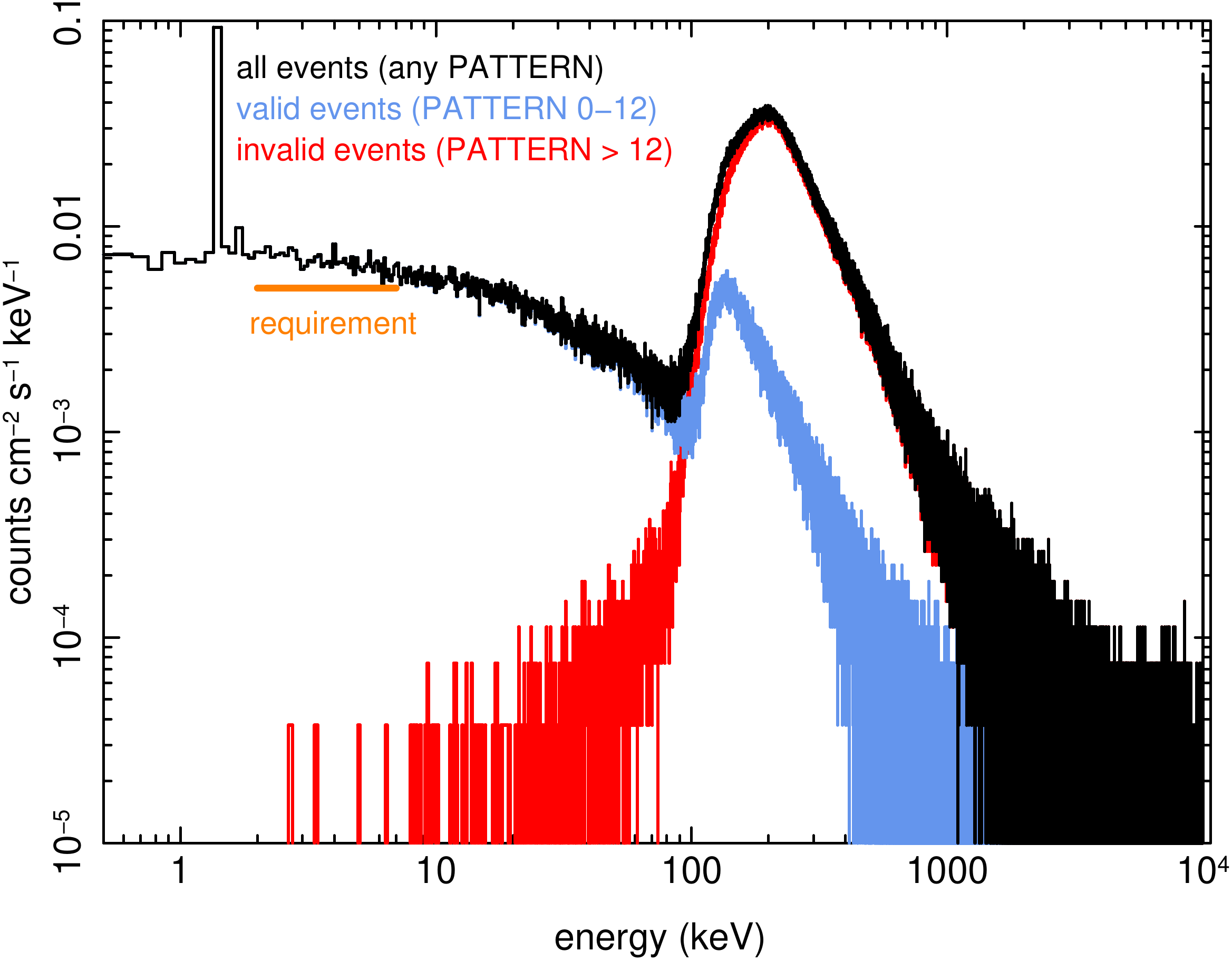}
\includegraphics[width=3.0in]{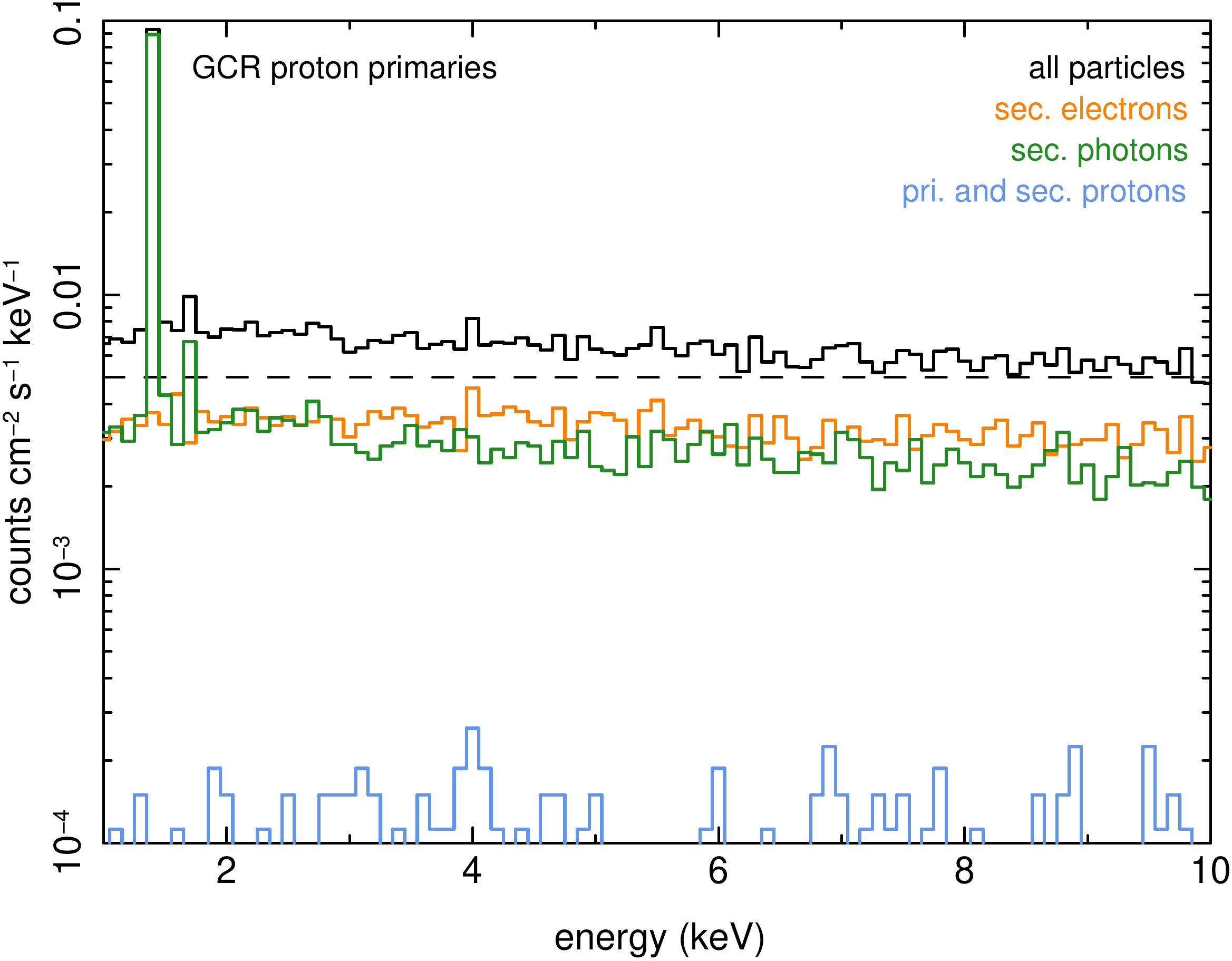}
\caption{Spectra of events produced by the Geant4 GCR proton primary simulations. (left) The spectrum over a wide energy band, showing pattern-based valid and invalid events separately. Valid events dominate by several orders of magnitude in the 2--7 keV band, while invalid events dominate above 100 keV, well outside the WFI sensitive band. (right) Spectrum in the 2--7 keV region, with the WFI unrejected background requirement of $5.5\times10^{-3}$ s$^{-1}$ cm$^{-2}$ keV$^{-1}$ plotted as a dashed line. Colored lines indicate what types of particles produce the detected signal for these events--primarily secondary electrons and photons produced in primary proton interactions with the WFI. The strong line near 1.5 keV is Al K$\alpha$, and the weaker line near 1.7 keV is Si K$\alpha$.}
\label{fig:event_spectra}
\end{center}
\end{figure}

Each image was searched for events using a local-maximum method similar to that employed on-board many X-ray imaging instruments like \xmm\ EPIC pn and \chandra\ ACIS. First an event threshold of 0.1~keV was applied, and pixels at or above this level were flagged as event candidates. Each candidate pixel was compared to the other pixels in its 3$\times$3 pixel neighborhood, and if it was a local maximum it was flagged as an event center. The 5$\times$5 neighborhood around each event center was then searched for pixels at or above the neighbor (or split) threshold, also set at 0.1~keV. The event pattern was assigned using EPIC pn rules\footnote{See the \xmm\ Science Analysis System User Guide, \url{https://xmm-tools.cosmos.esa.int/external/xmm_user_support/documentation/sas_usg/USG/SASUSG.html}},  including single-pixel events (PATTERN=0), doubles (PATTERN=1--4), triples (PATTERN=5--8), quadruples (PATTERN=9--12), and everything else (PATTERN=13). In particular, for all non-single-pixel events which have a 3$\times$3 neighbor above the neighbor threshold, the outer 5$\times$5 was also searched for pixels above the neighbor threshold. Double, triple, and quad patterns with at least one outer 5x5 pixel above the neighbor threshold were assigned PATTERN=13. In the remainder of this section, ``valid'' events are those with PATTERN$<$13, as these are indistinguishable from events produced by X-ray photons. The energy of the event is the summed energy of all pixels in the inner 3$\times$3 island that are above the neighbor threshold. Figure \ref{fig:event_spectra} shows the spectra of valid, invalid, and all events.

\begin{figure}[htpb]
\begin{center}
\includegraphics[width=4.0in]{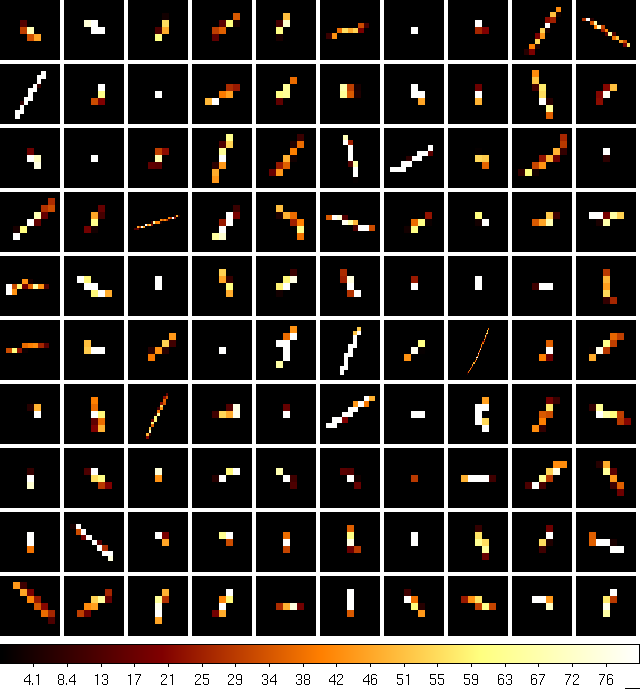}
\vspace{0.1in}
\caption{Images of a small sample of individual particle tracks in the WFI detector, with the color scale in keV. Pixels shown are equivalent to 130~$\mu$m WFI pixels, so the image sizes are not the same and scale with the size of the tracks.}
\label{fig:blob_pstamps}
\end{center}
\end{figure}

We identified particle tracks using image segmentation in each frame. Hereafter, a  ``particle track'' is defined as either (1) a spatially contiguous set of five or more pixels above the neighbor threshold, 0.1~keV; or (2) any contiguous set of pixels above 0.1~keV that includes at least one pixel over 15~keV. This latter energy is called the ``MIP threshold'', an energy above which the Athena mirrors have effectively zero efficiency, and thus all signal is assumed to be produced by cosmic ray minimum ionizing particles, or ``MIPs''. Detached diagonals are considered contiguous in this image segmentation. Each particle track was assigned an ID number to uniquely identify it in the full dataset. Examples of particle tracks are shown as postage stamps in Figure \ref{fig:blob_pstamps}. 

\begin{figure}[htpb]
\begin{center}
\includegraphics[width=4.0in]{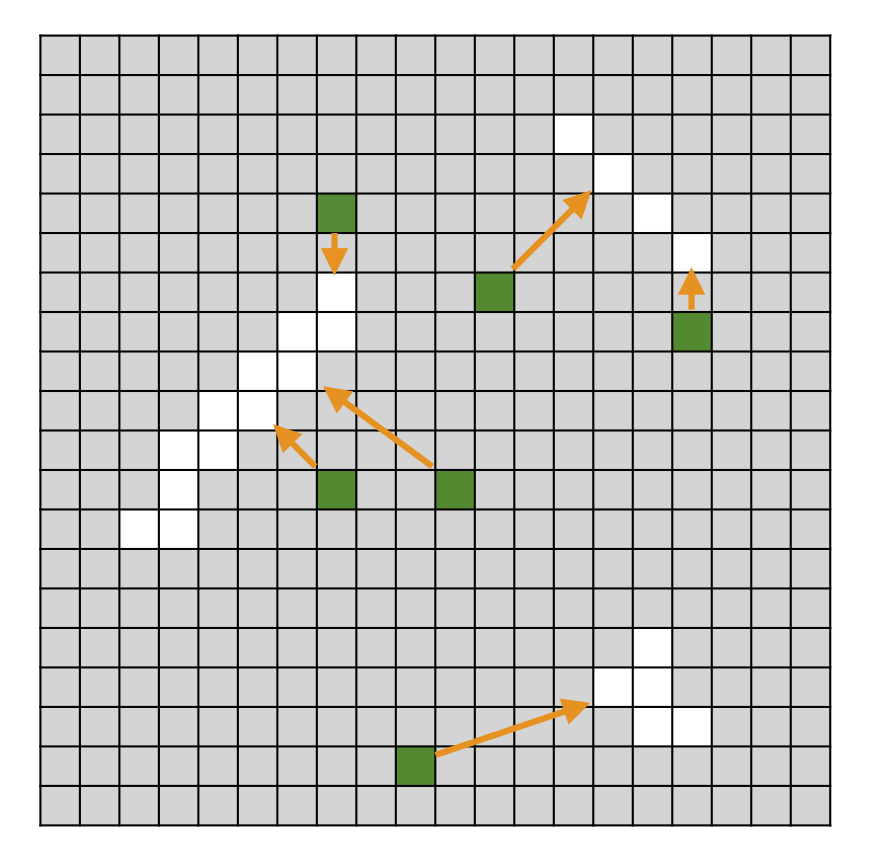}
\caption{Schematic of a frame containing particle tracks (white pixels) and valid events (green pixels). The image segmentation would identify three particle tracks in this frame. Orange arrows indicate the distance between each valid event, defined by the maximum pixel of the 3$\times$3 island, and the nearest particle track or MIP pixel.}
\label{fig:distance_scheme}
\end{center}
\end{figure}

Finally, in each frame, the distance between the central pixel of each event and the nearest pixel in a particle track was calculated. Many events fall on particle tracks, in which case this distance was set to zero. Valid events, however, are by definition unable to fall on a particle track pixel. Thus valid events and particle tracks are a mutually exclusive set of entities, despite the different methods used to identify them. A schematic diagram of this distance finding technique is shown in Figure \ref{fig:distance_scheme}.

To aid our analysis of the correlations between particle tracks and valid events, we assigned frames to ``cases'', namely:
\begin{itemize}
    \item Case A: frame contains only particle tracks.
    \item Case B: frame contains only valid events.
    \item Case C: frame contains both particle tracks and valid events.
    \item Case D: frame contains neither particle tracks nor valid events (empty frame).
\end{itemize}
This sorting was done for the single-primary frames as well as the 2-ms and 5-ms frames. Summary information about the fraction of frames and rates of particle tracks and valid events in each case is given in Table \ref{tab:frame_summary}.  

\begin{table}[htpb]
\begin{center}
\caption{Summary information for frame-by-frame analysis.}\label{tab:frame_summary}
\begin{tabular}{lrrr}\hline \hline
Type of Frame                   &  Single primary  &  2 msec           &  5 msec           \\
\hline
~no.~frames                      &  133,000,000     &  752,331          &  300,967          \\
~no.~frames with signal          &  935,504 (0.7\%) &  534,954 (71.1\%) &  287,361 (95.5\%) \\
~no.~frames with particle track  &  918,662 (0.7\%) &  530,114 (70.5\%) &  286,580 (95.2\%) \\
~no.~particle tracks per frame   &  0.0078          &  1.38             &  3.45             \\
\hline
\multicolumn{4}{l}{Case A (frame with only particle tracks)} \\
~fraction of all frames          &  0.68\%          &  67.8\%           &  87.3\%           \\
~fraction of frames with signal  &  97.3\%          &  95.3\%           &  91.4\%           \\
~no.~particle tracks per frame   &  1.12            &  1.94             &  3.57             \\
~fraction of valid events        &  ~$\cdots$~      &  ~$\cdots$~       &  ~$\cdots$~       \\
\hline
\multicolumn{4}{l}{Case B (frame with only valid events)} \\
~fraction of all frames          &  0.013\%         &  0.6\%            &  0.3\%            \\
~fraction of frames with signal  &  1.8\%           &  0.9\%            &  0.3\%            \\
~no.~particle tracks per frame   &  ~$\cdots$~      &  ~$\cdots$~       &  ~$\cdots$~       \\
~fraction of valid events        &  64.9\%          &  18.8\%           &  3.1\%            \\
\hline
\multicolumn{4}{l}{Case C (frame with both particle tracks and valid events)} \\
~fraction of all frames          &  0.007\%         &  2.7\%            &  7.9\%            \\
~fraction of frames with signal  &  0.9\%           &  3.8\%            &  8.3\%            \\
~no.~particle tracks per frame   &  1.89            &  2.51             &  4.20             \\
~fraction of valid events        &  35.1\%          &  81.2\%           &  96.9\%           \\
\hline
\multicolumn{4}{l}{Case D (frame with neither particle tracks nor valid events)} \\
~fraction of all frames          &  99.3\%          &  28.9\%           &  4.5\%            \\
~fraction of frames with signal  &  ~$\cdots$~      &  ~$\cdots$~       &  ~$\cdots$~       \\
~no.~particle tracks per frame   &  ~$\cdots$~      &  ~$\cdots$~       &  ~$\cdots$~       \\
~fraction of valid events        &  ~$\cdots$~      &  ~$\cdots$~       &  ~$\cdots$~       \\
\hline
\end{tabular}
\end{center}
\end{table}

\section{Spatial correlation between particle tracks and events}
\label{sect:analysis}

That valid events are spatially correlated with primary or secondary particle tracks from the same interacting cosmic ray was recognized early on in Geant4 simulations by the WFI Background Working Group\cite{wfibkg} and in the analysis of in-orbit \chandra, \swift, and \xmm\ data\cite{Grantetal2018,Bulbuletal2018,Bulbuletal2020}. This correlation can be exploited by masking around particle tracks and flagging valid events within a certain distance; such events can later be filtered in ground processing depending on the science goals of the observation. However, this masking also reduces the signal and thus the efficiency of the instrument. This optional, partial-veto method has been termed ``Self-Anti-Coincidence'' (SAC), since under this scheme the WFI detector acts as its own anti-coincidence detector (S.~Molendi, private communication and Ref.~\citenum{Miller20}). In Ref.~\citenum{Miller20}, we analyze the effects of SAC on different background reduction metrics, and explore the background improvement possible with enhanced, SAC-enabled post-processing algorithms.

\begin{figure}[htpb]
\begin{center}
\includegraphics[width=3.0in]{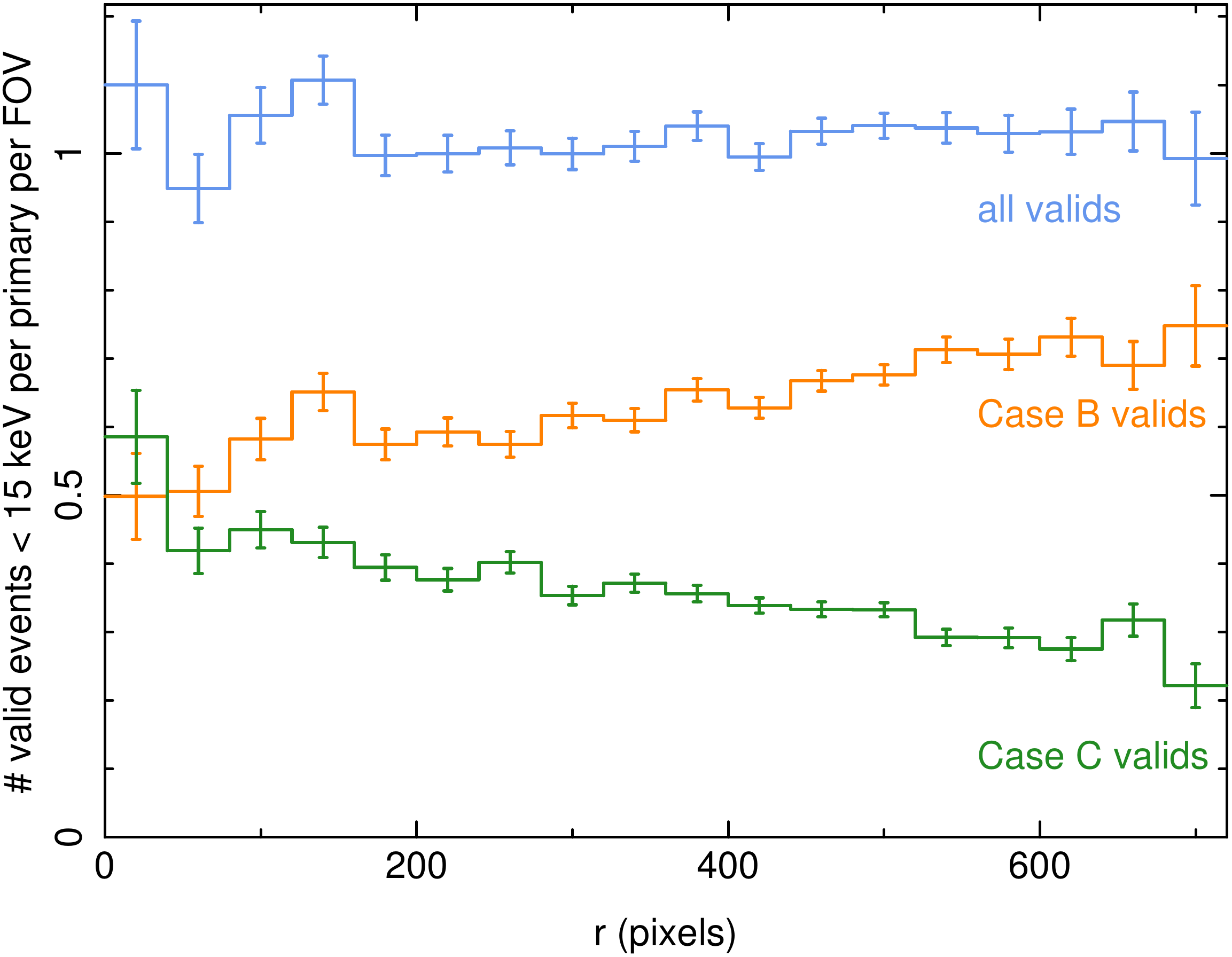}
\includegraphics[width=3.0in]{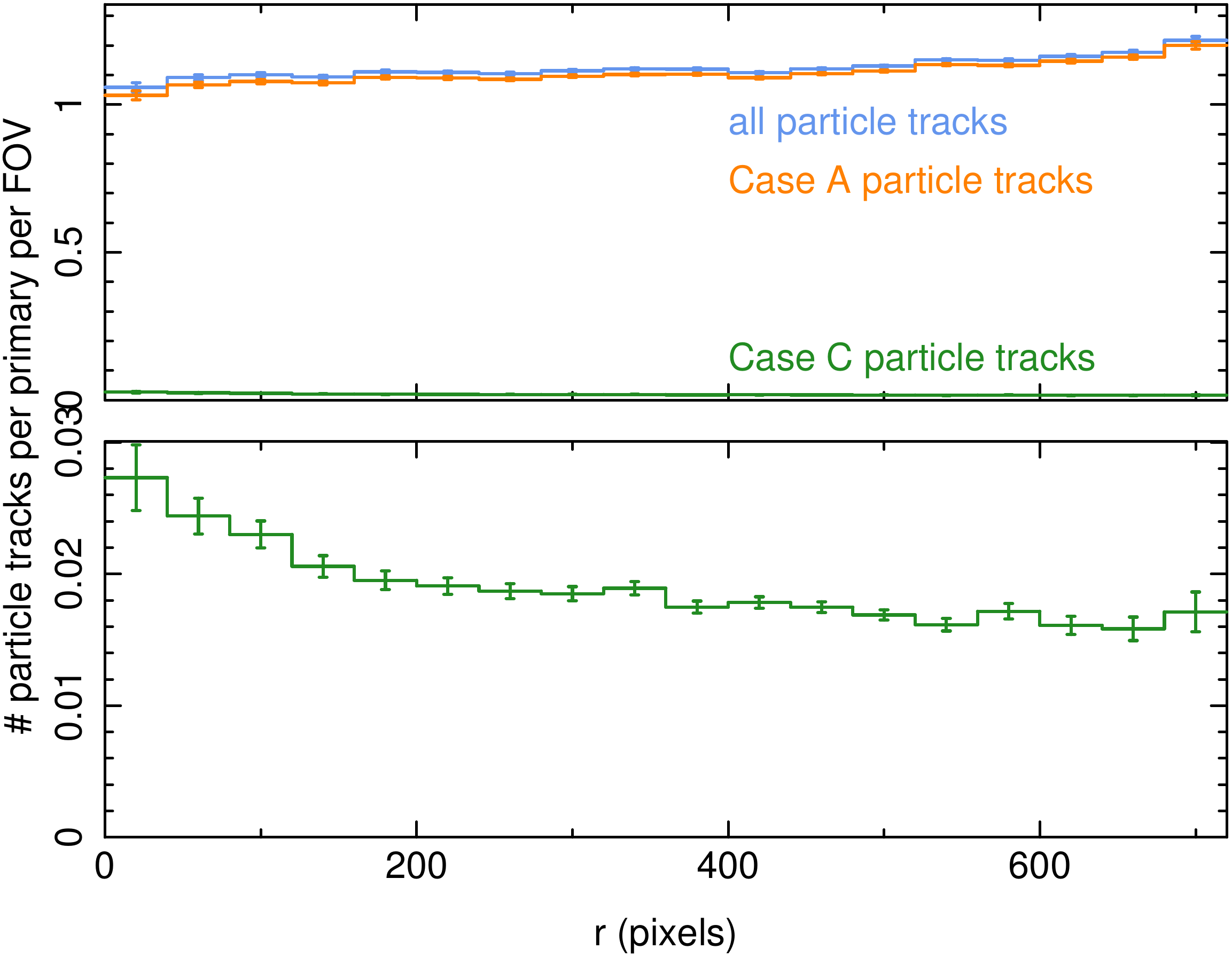}
\caption{Radial distribution of valid events (left) and particle tracks (right), normalized to the detector area. The lower panel of the right plot is a zoom-in to more clearly show the green points. The valid events overall have a flat distribution, however those valid events that accompany a particle track (Case C) are concentrated toward the center, and those that have no particle track (Case B) are more likely to be found near the edge. The particle tracks for those cases follow similar trends. This is expected, as a valid event detected near the edge is more likely to lose an accompanying particle track outside the field of view.}
\label{fig:raddist}
\end{center}
\end{figure}

Frames containing single cosmic ray primary particles are key to understanding the spatial correlation between particle tracks and valid events. The area-normalized radial distributions of valid events and particle tracks derived from these single-primary frames are shown in Figure \ref{fig:raddist}. While the valid events have a flat distribution overall, those that accompany particle tracks (Case C) are more likely to be found toward the center of the frame, and those that lack a particle track (Case B) are more likely near the edge. The particle tracks for those cases follow similar trends. This is expected, since a valid event detected near the edge is more likely to lose an accompanying particle track off the edge.  

\begin{figure}[p]
\begin{center}
\includegraphics[width=4.5in]{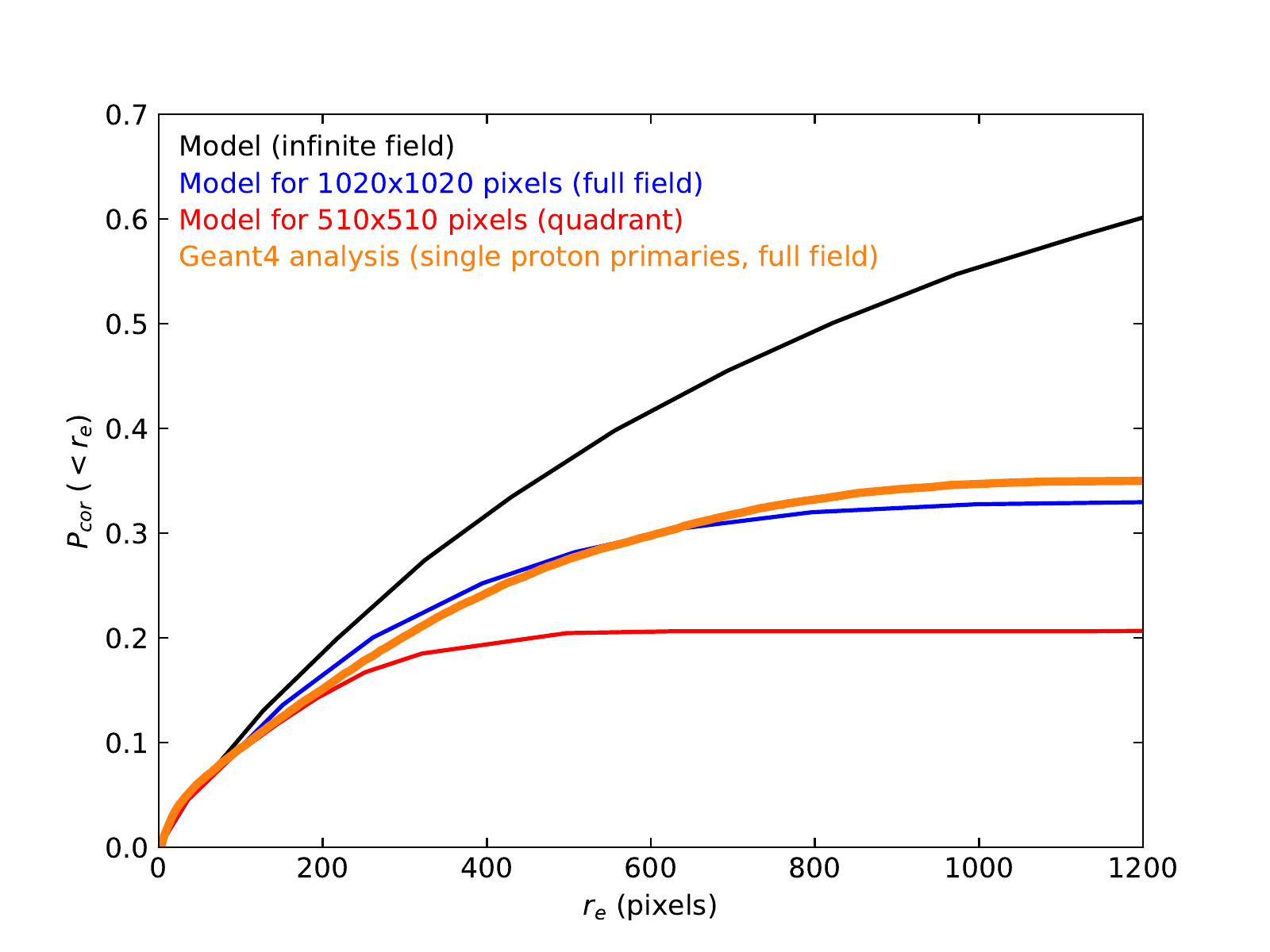}
\caption{Cumulative probability that a valid event falls within an exclusion radius $r_e$ of a particle track produced by the same primary. The orange line is derived from our single primary Geant4 simulation data. The other lines are from Ref.~\citenum{Miller20}.}
\label{fig:pcor}
\end{center}
\end{figure}

A useful metric to quantify this spatial correlation is the cumulative probability that a valid event falls within a certain radius of a particle track resulting from the same cosmic ray interaction.  We define this probability as \pcor$(<\,r_e)$, where $r_e$ is the ``exclusion radius'' to indicate its use in filtering unrejected background. A detailed analytic derivation of \pcor\, is presented in Ref.~\citenum{Miller20}, based on results from a previously published WFI Geant4 study\cite{wfibkg}. We determine \pcor\, empirically from our Geant4 results as the cumulative distribution of radius in pixels between all Case C valid events and the nearest pixel in a particle track.  To normalize \pcor\, to the full WFI field of view, we assume that Case B valid events have a corresponding particle track somewhere outside of the field. Thus we divide the distribution by the total number of valid events in Cases B and C. The resulting distribution is shown in Figure \ref{fig:pcor}, plotted with the analytic \pcor\, curves from Ref.~\citenum{Miller20}, with lines for an infinite plane (black), a full WFI field (blue), and a WFI quadrant (red). Our orange curve is consistent with the model for the full field, despite the very different methods used to derive the two. At the largest $r_e$, the correlation probability reaches 35\%. This is the maximum amount of effective background improvement we can achieve by using SAC; the other 65\% of valid events are produced in Case B primary interactions that do not also produce a particle track in the full field (see Table \ref{tab:frame_summary}).

\begin{figure}[p]
\begin{center}
\includegraphics[width=4.0in]{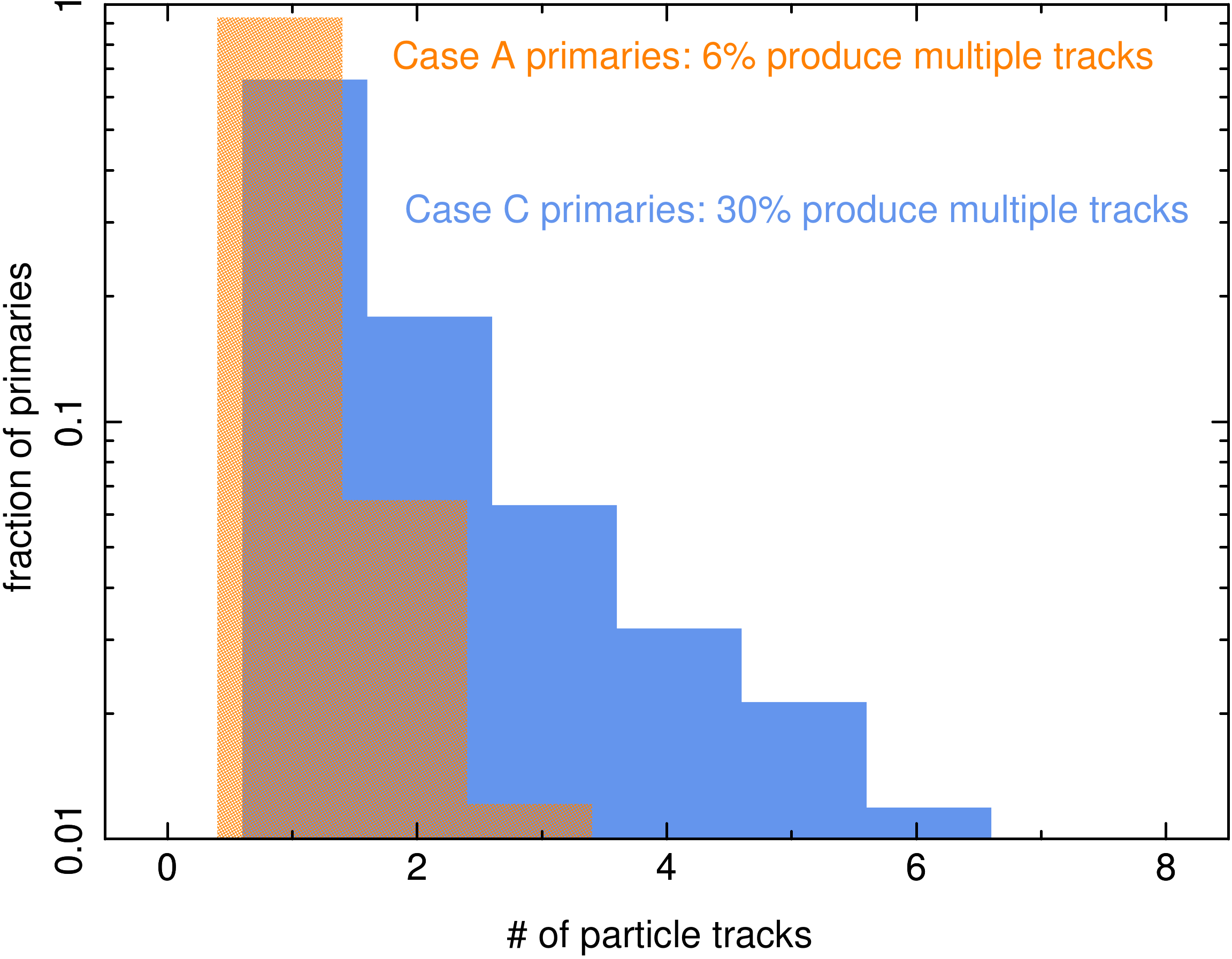}
\caption{Distribution of the number of particle tracks produced by primaries that do not also produce a valid event (Case A) and those that do (Case C). Valid events are more likely to be accompanied by a plurality of particle tracks. This can be used to identify frames that are likely to include valid events. Note that the histograms are shifted slightly along the X axis to improve clarity.}
\label{fig:havenot_numdist}
\end{center}
\end{figure}

In addition to a spatial correlation between particle tracks and valid events, we have found that proton primaries that produce valid events are much more likely to produce multiple particle tracks. This can also be seen from Table \ref{tab:frame_summary}, which shows that, among primaries that produce signal in the detector, Case A primaries produce on average 1.1 particle tracks, while Case C primaries produce 1.9 particle tracks. To further explore this, we plot in Figure \ref{fig:havenot_numdist} the distribution of particle track number for Case A and Case C primaries. Only 6\% of Case A primaries produce multiple particle tracks, whereas 30\% of Case C primaries do. Qualitatively, this makes sense; a primary interaction in the WFI structure can produce a shower of secondaries striking the detector, and these secondaries include both high-energy particles that produce tracks and lower energy photons and electrons that produce valid events. The number of independent particle tracks in a WFI frame contains some information about the likelihood of a valid event being present, and thus counting them could be a useful method to reduce the background. However, since this plurality occurs in 30\% of Case C primaries, and such primaries account for only 35\% of the valid events, no more than 10\% of the 2--7~keV background may be eliminated with this method. The potential gain is further reduced by the expectation of $\sim$3.5 particle tracks per 5-ms frame (see Table \ref{tab:frame_summary}). 

\section{Summary}
\label{sect:sum}

In this proceedings paper, we provide some highlights of our work to better understand and reduce the particle background on the Athena WFI.  Geant4 simulations of Galactic Cosmic Ray protons interacting with the WFI have been analyzed, producing lists of detected events and particle tracks.  Many of the detected events have pixel patterns and summed energies consistent with those from astrophysical X-ray sources. Spatial correlations between the events and particle tracks are being investigated as a discriminator between background and source events. Further details and analysis can be found in a more comprehensive work that is in preparation.\cite{Miller20}

\acknowledgments 
This work was done as part of the Athena WFI Background Working Group, a consortium including MPE, INAF/IASF-Milano, IAAT, Open University, MIT, SAO, and Stanford. We gratefully acknowledge support from NASA grant NNX17AB07G.

\bibliography{report} 
\bibliographystyle{spiebib} 

\end{document}